\documentclass[twoside,11pt]{article}

\usepackage{jmlr2e}
\usepackage{amsmath}
\usepackage{url}
\usepackage[usenames, dvipsnames]{color}

\newcommand{\X}{\mathbf{X}}

\newcommand{\Y}{\mathbf{Y}}

\newcommand{\W}{\mathbf{W}}

\newcommand{\R}{\mathbb{R}}

\jmlrheading{no}{2016}{pp}{sub}{pub}{Eemeli Lepp\"aaho, Muhammad Ammad-ud-din and Samuel Kaski}

\ShortHeadings{GFA: Group Factor Analysis}{Lepp\"aaho, Ammad-ud-din and Kaski}
\firstpageno{1}

\begin{document}

\title{GFA: Exploratory Analysis of Multiple Data Sources with Group Factor Analysis}

\author{\name Eemeli Lepp\"aaho \email eemeli.leppaaho@aalto.fi \AND
       \name Muhammad Ammad-ud-din  \email muhammad.ammad-ud-din@aalto.fi \AND
       \name Samuel Kaski \email samuel.kaski@aalto.fi \\
       \addr Helsinki Institute for Information Technology HIIT \\
       Department of Computer Science\\
       Aalto University\\
       P.O.Box 15400, FI-00076 Aalto, FINLAND
}

\editor{}

\maketitle

\begin{abstract}
The {\sc R} package GFA provides a full pipeline for factor analysis of multiple data sources that are represented as matrices with co-occurring samples. It allows learning dependencies between subsets of the data sources, decomposed into latent factors. The package also implements sparse priors for the factorization, providing interpretable biclusters of the multi-source data.
\end{abstract}

\section{Introduction}

The need for intelligent data integration and analysis methods is getting increasingly common.
We often encounter connected measurements from multiple data sources, possessing complex shared variation and structured noise.
Modelling the relationships within this kind of a data collection, and using them for prediction, is a challenging problem. Two existing {\sc R} packages, CCAGFA \citep{Klami15trnn} and CMF \citep{klami2014group}, provide good tools for this\footnote{Available at \url{https://cran.r-project.org/package=CCAGFA} and \url{https://cran.r-project.org/package=CMF} .}, but remain limited in their use for two reasons: the packages do not provide a complete data analysis pipeline, and they assume the structure present within data sources to be dense, that is, present in all the samples and the features.
This assumption is often unrealistic, and we provide package GFA to allow
sparse factorization of multi-source data, along with tools from data preprocessing to model interpretation.

Extending from the classical Bayesian matrix factorization work, \textit{group factor analysis} (GFA) \citep{Virtanen12aistats,Klami15trnn} assumes that data matrix $\Y^{(m)}\in\R^{N\times D_m}$ can be approximated with $\X \W^{(m)\top}$,
where the latent variable matrix $\X\in\R^{N\times K}$ describes the observed data in a lower dimensional space, and the projection matrix $\W^{(m)}\in\R^{D_m\times K}$ gives a mapping between the latent space and the data space of the $m$th source.
A key aspect of the model definition is a group sparse prior for the $\W$ matrices: the association of a data matrix to component $k$ is based on evidence in the data. Allowing only a subset of the components to explain a data matrix results in the posterior identifying relationships between any subset of the data sources.
The spike-and-slab sparsity priors used to achieve this are described by \cite{Bunte16bioinformatics}. Alternative priors aiming for similar goals have been presented by \cite{zhao2014bayesian} and made publicly available\footnote{\url{https://github.com/judyboon/BASS}}, though not containing any tools for preprocessing or model interpretation.

\section{The GFA Package}

GFA implements a Gibbs sampling scheme for inferring the posterior of group factor analysis, and it is available for {\sc R} at \url{https://cran.r-project.org/package=GFA}. It has been designed to contain the key elements needed for the data analysis task of factorization of multiple data sources.

\subsection{Model Options and Initialization}

The package includes a variety of model options for the user to choose from, depending on the application requirement. The default priors assume latent factors that are dense along the sample space, and group sparsity along the feature space \citep{Klami15trnn}. Using them, the posterior given data \verb|Y| can be inferred by: 
\begin{verbatim}
opts <- getDefaultOpts()
res <- gfa(Y,opts)
\end{verbatim}
Another main use of the package is biclustering of multiple data sources, which is done by using element-wise sparse priors for the matrices $\X$ and $\W$.
This approach has been presented in \cite{Bunte16bioinformatics}, and is usable via \verb|getDefaultOpts(bicluster=TRUE)|.

The input data \verb|Y| given to function \verb|gfa()| is a list of matrices (that is, data sources) that each have an equal number of rows. Alternatively, if there are several data sources with the same columns, it is possible to input a list with two elements: the sources with paired rows, and the sources with paired columns. Other important options include which parts of the data collection share the scale of each component or the strength of the residual noise (by default each data source for both), and whether the convergence of the sampling chain should be estimated (based on the Geweke diagnostic of its reconstruction). A full description of the possible options is given in the documentation of the function \verb|getDefaultOpts()|.

\subsection{Missing Value Prediction}

GFA can be used for prediction by simply including the missing matrix elements as \verb|NA|s. The model parameters are sampled based on the observed data only, and the missing values are predicted with the function \verb|reconstruction()|. If the data samples are acquired in a sequential manner, it is alternatively possible to i) infer the models parameters on the first batch, and ii) using fixed projections, quickly predict missing data sources of the second batch based on the observed data sources. 
Both the approaches use learned relationships between the data sources (shared components) to predict from one to other, and provide predictive distributions along with point estimates.

\section{Usage} 

In addition to inferring model parameters and predicting missing values, the data analysis pipeline typically needs additional steps presented here.

\subsection{Preprocessing}

The input data can be normalized in different ways with \verb|normalizeData(Y)|. As GFA aims to explain maximal variance with a compact set of parameters, and has no separate assumptions for the bias (mean) of features, centering the data is advised. Furthermore, the higher the variance of features (or data sources), the larger their effect to the model parameters. If the user considers feature variance as a proxy for it's importance, no scaling should thus be done. On the other hand, if the features (or data sources) are deemed approximately equally important, they should be scaled accordingly.


\subsection{Model Complexity Selection}

The priors of GFA are able to infer the model complexity. This is achieved by initializing the model with a high number of components $K$ (default $\min(N,\sum_m D_m)/2$) and pruning excessive components. As complexity selection is a difficult task in general, we recommend using any prior information of the data, if available, when selecting the initial $K$. Here it should be noted that the components can model (structured) noise in addition to the signal of interest, inducing a flexible noise model. If this is not desired, \verb|informativeNoisePrior()| can be used to specify how large proportion of variance is expected to be explained with the components; in effect causing weaker (structured noise) components to be explained by the simple residual noise. The package issues a warning if the automatic model complexity selection does not seem to work as intended, for example if no components are pruned.

\begin{figure}[t]
\centering
\includegraphics[width=0.49\textwidth,height=6cm,keepaspectratio]{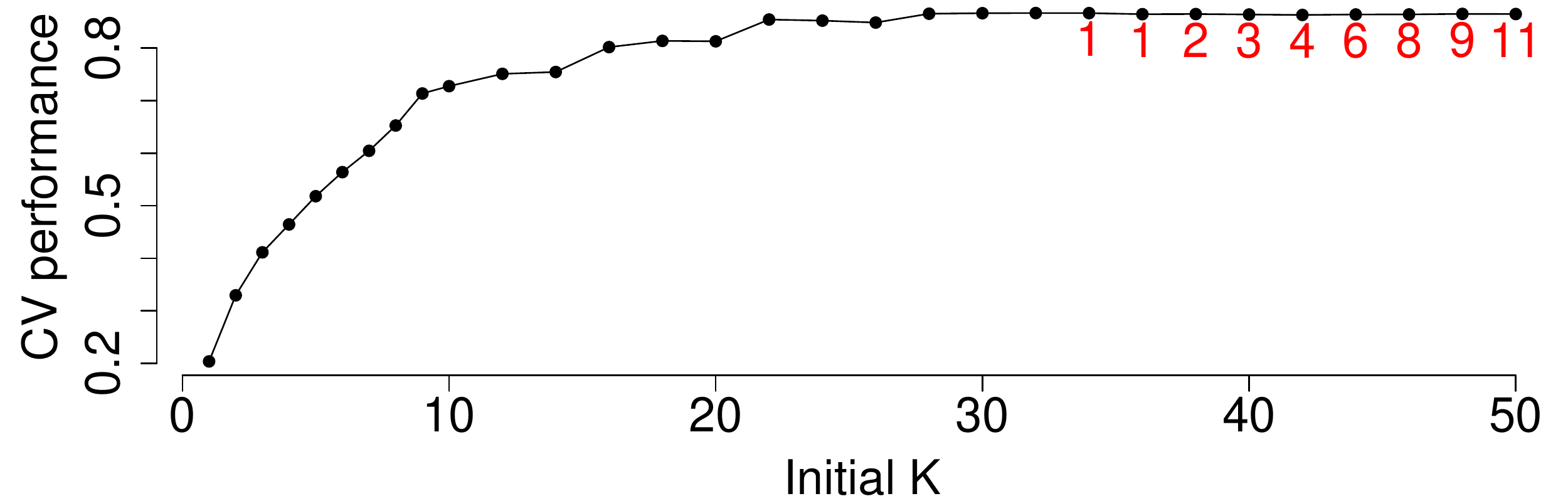}
\includegraphics[width=0.49\textwidth,height=6cm,keepaspectratio]{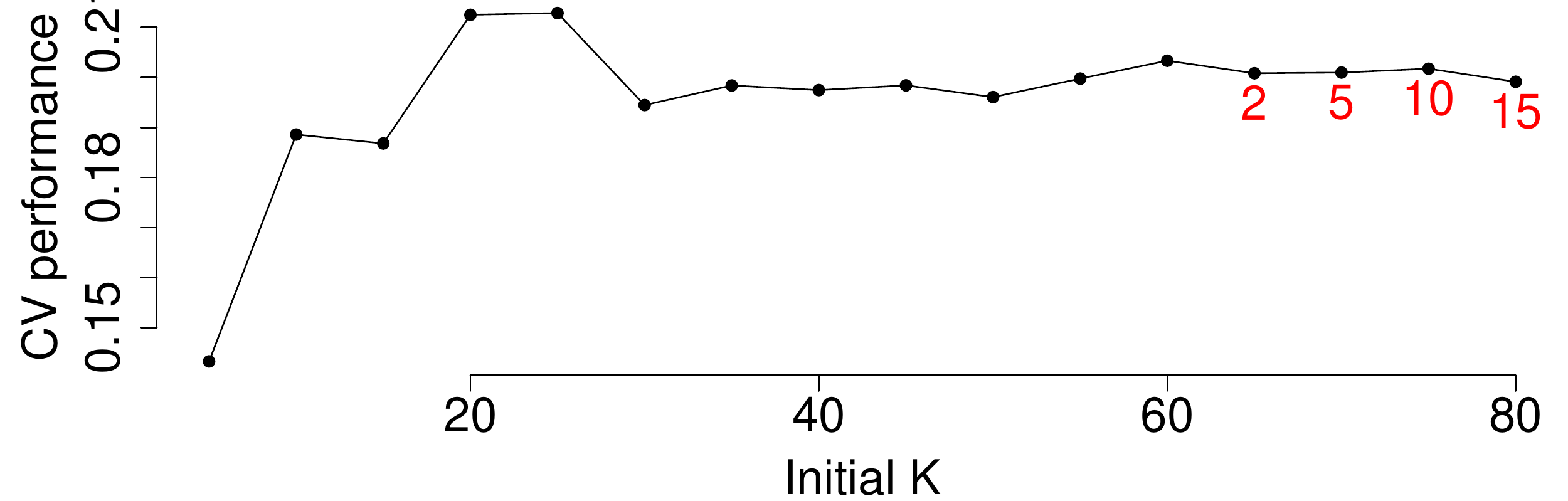}
\caption{Illustrating model complexity selection with GFA on synthetic data (left) and GDSC data set (right), showing the cross-validated predictive performance (Spearman correlation), and the number of empty components (red numbers) as a function of initial model complexity $K$. High initial $K$ allows the components to produce a flexible noise model, as opposed to low initial $K$.}
\label{fig}
\end{figure}

We demonstrate model complexity selection with two examples in Figure~\ref{fig}: a synthetic data set, and a cancer cell line data set by Genomics of Drug Sensitivity in Cancer project (GDSC)~\citep{yang2013genomics}. The complexity of the synthetic data set is overestimated somewhat, resulting in accurate predictions as long as initial $K$ is above the true $K$. With the GDSC data set, where the true model complexity is unknown, initial $K$ at 20 or 25 gives the best predictions, but the full model is not far behind.

\subsection{Interpreting the Model}

The inferred factorization explains dependencies in the data collection through $K$ components. The factorization can be visualized using function \verb|visualizeComponents()|, which shows the component activities in the data sources. Shared components imply predictability from a data source to another. The function can also show the full parameter matrices $\X$ and $\W$, as well as their effect, and the effect of individual components, in the data space.

\subsection{Robustness Analysis}

As in factor analysis in general, the factorization GFA aims to infer is unidentifiable, resulting in an extremely multi-modal posterior.\footnote{Unimodal posterior could be achieved in special cases, given limiting identifiability constraints.} Thus consideration is needed if a user wishes to analyze multiple factorizations (in this case sampling chains) jointly, as they are not likely to be identical. To the best of our knowledge, there are no sampling methods that would in practice converge to the true multi-modal posterior with high-dimensional data.
To this end, we provide function \verb|robustComponents()| which uses a correlation-based procedure to analyze which components occur frequently in different GFA sampling chains. This approach can be used to obtain robust factorizations for noisy data collections, where the components explain only small parts of the variation, resulting in uncertain estimates.

\subsection{Examples}

The package manual contains simplified examples demonstrating function usage. Besides this, we provide
the following three demonstrations:
 \begin{itemize}
 \setlength\itemsep{-0.5em}
 \item{demo(GFApipeline):} Simple illustration of the GFA pipeline.
\item{demo(GFAexample):} Elaborate illustration of the GFA model on simulated data.
\item{demo(GFAdream):} Replication of \cite{Bunte16bioinformatics} results (requiring data download, which is instructed).
 \end{itemize}


\section{Discussion}

GFA allows inferring the relationships between multiple co-occurring data matrices by reconstructing them with group sparse priors for the projection matrices.
It finds a decomposition with components that are specific to the data sources, shared between them all, or shared between any subset of them.
This kind of sparsity is beneficial for exploratory data analysis and integration. Our GFA package covers essential tools ranging from  preprocessing to modelling assumptions and from robustness analysis to interpreting the model.


\acks{This work was financially supported by the Academy of Finland (Finnish Center of Excellence in Computational Inference Research COIN; grants  295503 and 292337 to MA and SK). We acknowledge the computational resources provided by Aalto Science-IT project.}


%


\vskip 0.2in
\bibliography{gfa}

\end{document}